\documentstyle[aps]{revtex}
\begin{document}
\draft
\author{B.V.Ivanov}
\title{Algebraically special, expanding and twisting gravitational fields with
vanishing NUT parameter}
\address{Institute for Nuclear Research and Nuclear Energy\\
Tzarigradsko Shausse 72, Sofia 1784, Bulgaria}
\date{14 May 1999}
\maketitle

\begin{abstract}
The system of Einstein-Maxwell equations for fields mentioned in the title
is simplified. Known pure radiation solutions are systematized and new
solutions are given by separating the variables.
\end{abstract}


\section{Introduction}

There exist many papers dealing with algebraically special, expanding and
twisting vacuum, pure radiation or Einstein-Maxwell fields. An extensive
bibliography up to 1980 exists in [1]. Further results on Einstein-Maxwell
fields may be found in [2]. Pure radiation fields have been studied
extensively in [3-8]. Petrov type II fields are characterized by the
non-vanishing of the second Weyl scalar $\Psi _2$: 
\begin{equation}
\Psi _2=\left( m+iM\right) \rho ^3  \label{1}
\end{equation}
Here $m$ is the mass parameter, $M$ is the NUT parameter and $\rho $ is
given by 
\begin{equation}
\rho =-\frac 1{r+i\Sigma }  \label{2}
\end{equation}
where $r$ is the coordinate along the null congruence of geodesics and $%
\Sigma $ is the twist [1].

It has been noticed in different contexts that the condition $M=0$
simplifies the equations [2,4,5,7,9]. In the present paper we explore this
condition with the help of the method proposed in [3], putting the emphasis
on intrinsically time-dependent solutions. We clarify the general structure
of the solutions obtained by separation of the traditional variables $r,$ $%
u, $ $\zeta ,$ $\bar \zeta $ where $u$ is the retarded time and $\zeta ,$ $%
\bar \zeta $ span a two-dimensional surface. Known solutions are
systematized and new solutions are given.

Twisting gravitational fields with $M=0$ generalize the classes of
Robinson-Trautman (RT) [10] and Kerr-Schild (KS) fields [11] and are
physically realistic, their simplest representatives being the
Schwarzschild, Kerr and Wadia solutions.

In Sec. II the method of Stephani is applied to simplify the
Einstein-Maxwell equations. In Sec. III the main field equation is reduced
to a linear second order equation. Its general solution with separated
variables is found and studied in Sec. IV. Sec. V is devoted to solutions
linear in $u$, while in Sec. VI solutions with exponential behaviour are
discussed. Sec. VII contains some conclusions.

\section{The metric and field equations}

The standard form of the metric for expanding and twisting fields is [1]: 
\begin{equation}
ds^2=\frac{2d\zeta d\bar \zeta }{\rho \bar \rho P^2}-2\Omega \left[
dr+Wd\zeta +\bar Wd\bar \zeta +H\Omega \right]  \label{3}
\end{equation}

\[
\Omega =du+Ld\zeta +\bar Ld\bar \zeta 
\]

It is determined by the $r$-independent real functions $P,$ $m,$ $M$ and the
complex function $L$: 
\begin{equation}
2i\Sigma =P^2\left( \bar \partial L-\partial \bar L\right)  \label{4}
\end{equation}
\begin{equation}
W=\rho ^{-1}L_u+i\partial \Sigma  \label{5}
\end{equation}
\begin{equation}
H=-r\left( \ln P\right) _u-(mr+M\Sigma -\kappa \Phi _1^0\bar \Phi _1^0)\rho 
\bar \rho +\frac K2  \label{6}
\end{equation}
\begin{equation}
K=2P^2Re\left[ \partial \left( \bar \partial \ln P-\bar L_u\right) \right]
\label{7}
\end{equation}
where $\partial =\partial _\zeta -L\partial _u$ and $\kappa $ is the Newton
constant. The basic functions satisfy the following system of equations: 
\begin{equation}
\left( \partial -3L_u\right) \left( m+iM\right) =-2\kappa P^{-1}\Phi _1^0%
\bar \Phi _2^0  \label{8}
\end{equation}
\begin{equation}
P^{-3}M=Im\partial \partial \bar \partial \bar \partial V  \label{9}
\end{equation}
\begin{equation}
\frac{n^2}{2P^2}=-P\left[ P^{-3}\left( m+iM\right) \right] _u+P\left(
\partial \partial \bar \partial \bar \partial V\right) _u-\left( \partial
\partial V\right) _u\left( \bar \partial \bar \partial V\right) _u
\label{10}
\end{equation}
\begin{equation}
\left( \partial -2L_u\right) \Phi _1^0=0  \label{11}
\end{equation}
\begin{equation}
\left( \partial -L_u\right) \left( P^{-1}\Phi _2^0\right) +\left( P^{-2}\Phi
_1^0\right) _u=0  \label{12}
\end{equation}
where $V_u=P,$ $\Phi _1^0$ and $\Phi _2^0$ are the essential parts of the
Maxwell scalars 
\begin{equation}
\Phi _1=\rho ^2\Phi _1^0  \label{13}
\end{equation}
\begin{equation}
\Phi _2=\rho \Phi _2^0+\rho ^2P\left( 2\bar L_u-\bar \partial \right) \Phi
_1^0+2i\rho ^3P\left( \Sigma \bar L_u-\bar \partial \Sigma \right) \Phi _1^0
\label{14}
\end{equation}
and $n$ is either the energy density of pure radiation (then $\Phi _1=\Phi
_2=0$ ) or is given in the Einstein-Maxwell case by 
\begin{equation}
n^2=2\kappa \Phi _2^0\bar \Phi _2^0  \label{15}
\end{equation}
Vacuum solutions have $\Phi _1=\Phi _2=n=0$.

The form of the metric is preserved by certain coordinate transformations,
one of which is the change of the retarded time 
\begin{equation}
u^{\prime }=F\left( u,\zeta ,\bar \zeta \right)  \label{16}
\end{equation}
\begin{equation}
P^{\prime }=F_u^{-1}P  \label{17}
\end{equation}
\begin{equation}
L^{\prime }=F_uL-F_\zeta  \label{18}
\end{equation}
\begin{equation}
\Sigma ^{\prime }=F_u^{-1}\Sigma  \label{19}
\end{equation}
\begin{equation}
\left( m+iM\right) ^{\prime }=F_u^{-3}\left( m+iM\right)  \label{20}
\end{equation}

Let us apply now to the system (8-12) the method of Stephani [3]. It works
when either $\Phi _1^0$ or $\Phi _2^0$ vanishes. Then equation (8) is solved
by 
\begin{equation}
m+iM=\phi _u^3  \label{21}
\end{equation}
The complex field $\phi $ is invariant under equation (16). It satisfies $%
\partial \phi =0$, which stated differently gives 
\begin{equation}
L=\frac{\phi _\zeta }{\phi _u}  \label{22}
\end{equation}
When $M=0$ we can use equation (16) to transform $m$ to a positive or
negative constant $m_0$ so that 
\begin{equation}
\phi =m_0\left[ u+iq\left( \zeta ,\bar \zeta \right) \right]  \label{23}
\end{equation}
\begin{equation}
L=iq_\zeta  \label{24}
\end{equation}
Obviously $L_u=0$. This gauge differs from the most popular Kerr's gauge
when $P_u=0$, but is very suitable when the NUT parameter vanishes. Eqs. (9)
and (10) become 
\begin{equation}
\partial \partial \bar \partial \bar \partial V=\bar \partial \bar \partial
\partial \partial V  \label{25}
\end{equation}
\begin{equation}
\frac{n^2}{2P^2}=3m_0P^{-3}P_u+P\partial \partial \bar \partial \bar \partial
P-\partial \partial P\ \bar \partial \bar \partial P  \label{26}
\end{equation}
with $\partial =\partial _\zeta -iq_\zeta \partial _u$. When $P_u\neq 0,$ $%
n^2$ can be made positive by the choice of $m_0$ at least for some region of
spacetime [1,4,9].

If $\Phi _1^0$ $=0$ Eq. (12) gives 
\begin{equation}
\Phi _2^0=PG\left( \phi ,\bar \zeta \right)  \label{27}
\end{equation}
where $G$ is an arbitrary function. If $\Phi _2^0=0$ Eqs. (11) and (12) give 
\begin{equation}
\Phi _1^0=\bar C(\bar \zeta )\exp \left[ 2i\int \left( \ln P\right)
_uq_\zeta d\zeta \right]  \label{28}
\end{equation}
with $C\left( \zeta \right) $ being an arbitrary analytic function. This is
a particular case of Theorem 26.2 from [1]. The expressions for the other
functions in the metric simplify: 
\begin{equation}
\Sigma =P^2Q  \label{29}
\end{equation}
\begin{equation}
W=i\partial \left( P^2Q\right)  \label{30}
\end{equation}
\begin{equation}
K=P^2\left( \bar \partial \partial +\partial \bar \partial \right) \ln P
\label{31}
\end{equation}
\begin{equation}
H=-r\left( \ln P\right) _u-\left( m_0r-\kappa \Phi _1^0\bar \Phi _1^0\right)
\rho \bar \rho +\frac K2  \label{32}
\end{equation}
We have introduced the real, invariant under (16), function 
\begin{equation}
Q\left( \zeta ,\bar \zeta \right) =q_{\zeta \bar \zeta }  \label{33}
\end{equation}
The Weyl scalars [12] contain a lot of terms with $L_u$ and also simplify in
the gauge $L_u=0$. We quote only the leading terms: 
\begin{equation}
\Psi _2=m_0\rho ^3  \label{34}
\end{equation}
\begin{equation}
\Psi _3=-\rho ^2P^3\partial I+O\left( \rho ^3\right)  \label{35}
\end{equation}
\begin{equation}
\Psi _4=\rho P^2I_u+O\left( \rho ^2\right)  \label{36}
\end{equation}
\begin{equation}
I=P^{-1}\bar \partial \bar \partial P  \label{37}
\end{equation}

What remains to be solved is the couple of equations (25) and (26) for $P$
and $q$. For pure radiation Eq. (26) is just an inequality.

When $m+iM=0$ (Petrov types III and N) Eq.(8) holds identically, but still a
potential $\phi $ may be introduced with the property $\partial \phi =0$ and
we can study the subclass of solutions satisfying Eq. (23) (with $m_0=1)$.
Then in all other equations we can set $m_0=0$.

\section{The field equation (25)}

Eq. (25) is a linear equation of fourth order with respect to $V$. However,
in the gauge $L_u=0$ it becomes a linear equation of second order for $P$.
This can be established with the help of the commutator 
\begin{equation}
\left[ \partial ,\bar \partial \right] =2iQ\partial _u  \label{38}
\end{equation}
Then Eq. (25) becomes 
\begin{equation}
Q_{\zeta \bar \zeta }P+2Q_{\bar \zeta }P_\zeta +2Q_\zeta P_{\bar \zeta
}+4QP_{\zeta \bar \zeta }+2i\left( q_{\bar \zeta }Q_\zeta -q_\zeta Q_{\bar 
\zeta }\right) P_u+4iQ\left( q_{\bar \zeta }P_{\zeta u}-q_\zeta P_{\bar \zeta
u}\right) +4Qq_\zeta q_{\bar \zeta }P_{uu}=0  \label{39}
\end{equation}
This equation characterizes the twisting solutions because when $\Sigma =0$
it is trivial. It is non-linear in $q$ except for time-independent solutions
when only the first four terms remain. In this case Eq. (26) turns into the
equation for type III RT solutions, containing only $P$. It is logical to
solve first Eq. (26) for $P$ and plug the result in Eq. (39) to find $q$. In
the present paper time-dependent solutions will be discussed mainly. In this
case Eq. (39) should be solved for $P$ when $q$ is given and the result
placed in Eq. (26) to find $n^2$ for pure radiation solutions.

Similar equations have been derived for time-independent fields in [1] (see
Eq. (25.46)) and [2] (Eq. (5.12)). For time-dependent fields and concrete
expressions for $q$ such equations can be found in [7,9]. A linear equation
of second order for $\left( -2\Sigma \right) ^{1/2}$ is found in [5] in
terms of Cauchy-Riemann structures admitting Lie groups of symmetries and
used in [4-6] to obtain pure radiation solutions.

If $P$ and $q$ depend on $\zeta $ and $\bar \zeta $ via a single function $%
\alpha \left( \zeta ,\bar \zeta \right) $ the terms in brackets in Eq. (39)
cancel and it yields 
\begin{equation}
\left( Q_{\alpha \alpha }\alpha _\zeta \alpha _{\bar \zeta }+Q_\alpha \alpha
_{\zeta \bar \zeta }\right) P+4\left( Q_\alpha \alpha _\zeta \alpha _{\bar 
\zeta }+Q\alpha _{\zeta \bar \zeta }\right) P_\alpha +4Q\alpha _\zeta \alpha
_{\bar \zeta }\left( P_{\alpha \alpha }+q_\alpha ^2P_{uu}\right) =0
\label{40}
\end{equation}
Let $\zeta =\frac 1{\sqrt{2}}\left( x+iy\right) $. Suppose that $\alpha =x$
and $\partial _y$ is a Killing vector. Then equation (40) becomes 
\begin{equation}
q_x^2QP_{uu}+QP_{xx}+Q_xP_x+\frac 14Q_{xx}P=0  \label{41}
\end{equation}
and $Q=\frac 12q_{xx}$. In the case of axial symmetry $\alpha =\sigma =\zeta 
\bar \zeta $ and Eq. (40) reads 
\begin{equation}
q_\sigma ^2P_{uu}+P_{\sigma \sigma }+\frac{\left( \sigma Q\right) _\sigma }{%
\sigma Q}P_\sigma +\frac{\left( \sigma Q_\sigma \right) _\sigma }{4\sigma Q}%
P=0  \label{42}
\end{equation}
\begin{equation}
Q=\left( \sigma q_\sigma \right) _\sigma  \label{43}
\end{equation}
The introduction of $z=\ln \sigma $ as a variable greatly simplifies Eqs.
(42) and (43) making them an analog of Eq. (41): 
\begin{equation}
q_z^2QP_{uu}+QP_{zz}+Q_zP_z+\frac 14Q_{zz}P=0  \label{44}
\end{equation}
\begin{equation}
Q=e^{-z}q_{zz}  \label{45}
\end{equation}
Thus every solution with y-symmetry has an axisymmetric mirror with the same 
$P$ and $Q$ but as functions of $z$.

\section{Separation of variables}

Let us search for twisting solutions by separating the variables $P=p\left(
\zeta ,\bar \zeta \right) f\left( u\right) $. Eq.(39) transforms into 
\begin{equation}
f_{uu}+Af_u+Cf=0  \label{46}
\end{equation}
\begin{equation}
A=\frac i2\left[ \frac{Q_\zeta }{q_\zeta Q}-\frac{Q_{\bar \zeta }}{q_{\bar 
\zeta }Q}+2\left( \frac{p_\zeta }{q_\zeta p}-\frac{p_{\bar \zeta }}{q_{\bar 
\zeta }p}\right) \right]   \label{47}
\end{equation}
\begin{equation}
C=\left( 4q_\zeta q_{\bar \zeta }Qp\right) ^{-1}\left( Q_{\zeta \bar \zeta
}p+2Q_{\bar \zeta }p_\zeta +2Q_\zeta p_{\bar \zeta }+4Qp_{\zeta \bar \zeta
}\right)   \label{48}
\end{equation}
Eq. (46) is a well-known linear equation. It requires that $A$ and $C$ must
be constant and possesses three types of solutions according to the sign of $%
\lambda ^2=4C-A^2$: 
\begin{equation}
f=e^{-\frac{Au}2}\left( C_1e^{\frac{\beta u}2}+C_2e^{-\frac{\beta u}2%
}\right)   \label{49}
\end{equation}
\begin{equation}
f=e^{-\frac{Au}2}\left( C_1\sin \frac{\lambda u}2+C_2\cos \frac{\lambda u}2%
\right)   \label{50}
\end{equation}
\begin{equation}
f=e^{-\frac{Au}2}\left( C_1u+C_2\right)   \label{51}
\end{equation}
where $\beta ^2\equiv -\lambda ^2>0,$ $\lambda ^2>0$ and $\lambda =0$
respectively. Notice the common exponential factor if $A\neq 0$. It is quite
interesting that damping exponential behaviour is generic for RT solutions,
but it stems from the analog of equation (26) in the non-twisting case
[13,14]. When Eq. (40) holds, $A$ vanishes. There are still three types of
solutions, one of them with exponential behaviour. When there is y-symmetry
Eqs. (46) and (41) become 
\begin{equation}
f_{uu}+Cf=0  \label{52}
\end{equation}
\begin{equation}
Qp_{xx}+Q_xp_x+\left( \frac 14Q_{xx}-Cq_x^2Q\right) p=0  \label{53}
\end{equation}
In the case of axial symmetry equation (52) still holds, but equation (53)
is replaced either by the same equation with $x$ changed to $z=\ln \sigma $
or by 
\begin{equation}
p_{\sigma \sigma }+\left( \ln \sigma Q\right) _\sigma p_\sigma +\left[ \frac{%
\left( \sigma Q_\sigma \right) _\sigma }{4\sigma Q}-Cq_\sigma ^2\right] p=0
\label{54}
\end{equation}

Let us discuss next Eq. (26) for two of the types of $u$ -behaviour allowed
by Eq. (39). Suppose first that $P=pe^{cu}$. Introduce the function $%
B=pe^{icq}$. Then Eq. (26) reads 
\begin{equation}
n^2=6m_0c+2p^2\left( \psi e^{-4icq}\right) _\zeta e^{2icq+4cu}  \label{55}
\end{equation}
\begin{equation}
\psi =BB_{\zeta \bar \zeta \bar \zeta }-B_\zeta B_{\bar \zeta \bar \zeta }
\label{56}
\end{equation}
When $m_0=0$ the solution is of type III only when $\psi \neq 0$ because $%
\partial I=\psi /B^2$. When $m_0\neq 0$ and $\psi =0$ the solution is of KS
type because $\partial I=I_\zeta =0$ and $I$ can be nullified by a
coordinate transformation $\bar \zeta ^{\prime }=\bar g\left( \bar \zeta
\right) $. This means $\partial \partial P=0$ (see Eq. (37)), which can be
lifted to $\partial \partial V=0$. The latter is exactly the Kerr-Schild
condition [1,3].

Suppose next that $P=pu$. When there is y-symmetry Eq. (26) yields 
\begin{equation}
n^2=\frac{6m_0}u+\frac 12p^2\left( pp_{xxxx}-p_{xx}^2\right) u^4-2p^2\left(
q_xp_x+Qp\right) ^2u^2  \label{57}
\end{equation}
When there is axial symmetry Eq. (26) changes into 
\begin{equation}
n^2=\frac{6m_0}u+2p^2\left[ 2pp_{\sigma \sigma }+4\sigma pp_{\sigma \sigma
\sigma }+\sigma ^2\left( pp_{\sigma \sigma \sigma \sigma }-p_{\sigma \sigma
}^2\right) \right] u^4-2\sigma ^2\left[ \left( q_\sigma p^2\right) _\sigma
\right] ^2u^2  \label{58}
\end{equation}

Finally, let us perform the separation of variables in the Maxwell
equations. Eq. (28) simplifies 
\begin{equation}
\Phi _1^0=\bar C\left( \bar \zeta \right) \exp \left[ 2iq\left( \ln f\right)
_u\right]  \label{59}
\end{equation}
When $f=u$ and $f=e^{cu}$ this becomes respectively 
\begin{equation}
\Phi _1^0=\bar C\left( \bar \zeta \right) e^{\frac{2iq}u}  \label{60}
\end{equation}
\begin{equation}
\Phi _1^0=\bar C\left( \bar \zeta \right) e^{2icq}  \label{61}
\end{equation}

Eq. (12) or (27) separates as follows 
\begin{equation}
P^{-1}\Phi _2^0=\bar N\left( \bar \zeta \right) e^{a\left( u+iq\right) }
\label{62}
\end{equation}
$N\left( \zeta \right) $ being another arbitrary analytic function and $a$
is a constant different from $c$ in general. The energy density of a null
Maxwell field (15) becomes 
\begin{equation}
\frac{n^2}{2P^2}=\kappa \mid N\left( \zeta \right) \mid ^2e^{2au}  \label{63}
\end{equation}
Obviously polynomial dependence of $P$ in Eq. (26) is not allowed by (63).
Consequently, null Maxwell fields can't induce solutions with $P=pu$. It is
seen also from Eqs. (55), (57) and (58) that vacuum solutions of type II are
not possible because the different terms cannot cancel each other.

\section{Solutions linear in $u$}

These are solutions with $C=0$ in Eq. (46) or Eqs. (52), (53) and (54). Then
these equations are identical to the equations for a $u$ -independent $P$.
Therefore we can use any time-independent solution of Eqs. (25) and (26)
found in the literature which has $L_u=0$ and axial or y-symmetry. We can
also use time-independent solutions of Eq. (53) or Eq. (54) which are not
solutions of Eq. (26).

For example, let us find polynomial solutions with y-symmetry which
capitalize on the only known vacuum RT solution for type III fields [1,10].
If we take $L=ix^{b-1},$ $b\neq 1$ then Eq. (53) becomes the Euler equation 
\begin{equation}
x^2p_{xx}+\left( b-2\right) xp_x+\frac 14\left( b-2\right) \left( b-3\right)
p=0  \label{64}
\end{equation}
and its solutions are 
\begin{equation}
p=\mid x\mid ^a  \label{65}
\end{equation}
\begin{equation}
p=\ln \mid x\mid  \label{66}
\end{equation}
\begin{equation}
p=\mid x\mid ^{\frac{3-b}2}\left[ C_1\sin \left( \mu \ln \mid x\mid \right)
+C_2\cos \left( \mu \ln \mid x\mid \right) \right]  \label{67}
\end{equation}
when $b<3,$ $b=3$ and $b>3$ respectively. Here $a=\frac 12\left( 3-b\pm 
\sqrt{3-b}\right) $ and $\mu =\frac 12\sqrt{b-3}$. The first of these
solutions has been obtained essentially in [4] by a rather intricate
procedure. Plugging Eq. (65) and $q_x=x^{b-1}$ into Eq. (57) yields the
inequality to be satisfied: 
\begin{equation}
\frac{12m_0}u\mid x\mid ^{2\left( b-1\pm \sqrt{3-b}\right) }+2a\left(
a-1\right) \left( 3-2a\right) u^4-\left( 2-\sqrt{3-b}\right) ^2\mid x\mid
^{2b}u^2\geq 0  \label{68}
\end{equation}
Throughout the paper we consider $u>0$. The first term is always positive
(we accept that $m_0>0$ unless stated otherwise), the third is always
negative, while the second changes sign and is positive for $a<0$ and $%
1<a<3/2$. The third term has negative poles in $x$ for any $b\neq 0$ and $%
b\neq -1$ which cannot be compensated by the second term. Hence, pure
radiation solutions of type III are not allowed and the first term should be
present to compensate the negative poles with positive ones. This is
achieved when $b<2$. The plus sign in the formula for $a$ must be taken in
both cases. For growing $u$, however, the first term diminishes to zero.
Thus the energy density is positive for $b<2$ and small enough $u$. When $%
b=0 $ the third term is $x$-independent but the second is negative,
excluding again a type III solution. Type II solution still exists. Finally,
when $b=-1 $ the third term vanishes but the second is still non-positive
because $a=1$ or $a=3$. When $a=1$ only the first term remains and $n$ is $x$%
-independent, positive and bounded. This is a KS solution because $\partial
\partial V=-xq_x^2/2$ and can be turned into zero by a $u$-independent
transformation of $V$.

The second solution, given by Eq. (66), has no free parameters and leads to
incurable negative poles and logarithmic singularities in $n$ and is
therefore unphysical. As a last comment , we may start with $p=x^{b-1}$ in
Eq. (53) and obtain an Euler equation for $Q$ with the corresponding three
types of solutions.

The simplest solution of Eq. (53) is probably when $P$ is $u$-dependent
only, i.e. $p=p_0=const$. Then $Q_{xx}=0$ and $P=p_0u$, $L=\frac i{\sqrt{2}}%
x\left( c_1x+2c_2\right) $ where $c_i$ are arbitrary constants and 
\begin{equation}
n^2=\frac{6m_0}u-2p_0^4\left( c_1x+c_2\right) ^2u^6  \label{69}
\end{equation}
The energy density is positive for any $x$ when $c_1=0$ and $u$ is small
enough. This is not a KS field.

Another simple solution is obtained when $p=e^{ax}$, $a$ being some
constant. Then Eq. (53) has constant coefficients and one of the solutions
is $q_x=\exp \left( -2ax\right) $. The energy density is positive, $%
n^2=6m_0/u$ and $I=a^2/2$. Consequently the solution is of type II and is
equivalent to a KS solution.

Let us investigate next the axisymmetric solutions. As has been mentioned in
the previous sections, Eq. (53) has a counterpart for axisymmetric fields
with the same $Q$ and $P$ but with $x$ replaced by $z$. However, $q$ and $n$
are different, being given by Eqs. (45) and (58) instead of $q_{xx}=2Q$ and
Eq. (57). The solution with $p=p_0$ has a mirror solution 
\begin{equation}
L=i\bar \zeta \left( c_3\ln \sigma +c_4\right)  \label{70}
\end{equation}
\begin{equation}
n^2=\frac{6m_0}u-2c_3^2p_0^4u^6  \label{71}
\end{equation}
The energy density is positive for small $u$. When $c_3=0$ it is positive
everywhere and the solution is a KS field.

Another solution has $q=\sigma ,$ $L=i\bar \zeta ,$ $Q=1$. Then Eq. (54)
yields 
\begin{equation}
p_{\sigma \sigma }+\frac 1\sigma p_\sigma =0  \label{72}
\end{equation}
and $p=c_5\ln \sigma +c_6$. It was found by a different method in [9]. When $%
c_5=0$ it coincides with the previous solution with $c_3=0$.

A well-known solution is given by $p=1+\sigma /2$. Eq. (54) becomes linear
with respect to $Q$: 
\begin{equation}
\left( 2+\sigma \right) \sigma Q_{\sigma \sigma }+\left( 2+5\sigma \right)
Q_\sigma +4Q=0  \label{73}
\end{equation}
and its solution is the hypergeometric function $F\left( 2,2,1,-\sigma
/2\right) $. It degenerates into a rational function for these values of its
parameters and we have 
\begin{equation}
Q=\frac{4k\left( 2-\sigma \right) }{\left( 2+\sigma \right) ^3}  \label{74}
\end{equation}
\begin{equation}
L=-\frac{ik\bar \zeta }{\left( 1+\sigma /2\right) ^2}  \label{75}
\end{equation}
where $k$ is an arbitrary constant, characterizing the magnitude of the
twist (rotation). This is the Kramer solution [1,7,8,9] which is of KS type
and has $n^2=6m_0/u$. It represents a radiating Kerr metric. This is seen
best in the gauge $P_u=0$.

It must be stressed that any solution of the type $P=pu$ can be transformed
from the $L_u=0$ gauge to the $P_u=0$ gauge by choosing $F=u^2/2$ in Eq.
(16). Then $P^{\prime }=p,$ $L^{\prime }=\left( 2u\right) ^{1/2}L$ $,$ $%
m^{\prime }=\left( 2u\right) ^{-3/2}m$ which transfers the $u$-dependence to
the mass parameter and the twist.

A generalization of the Kramer solution was undertaken in [7] for the same $%
L $ and the ansatz $P=A\left( \sigma ,u\right) \left( 1+\sigma /2\right) $.
It was shown that $A$ satisfies a second order linear equation. The results
of this paper guarantee that $P$ also satisfies a second order linear
equation.

The mirror of the solution with $p=e^{ax}$ described above is $p=e^{az}$ and 
$Q=k\left( 1-2a\right) e^{-2az},$ $a\neq 1/2$. Hence 
\begin{equation}
P=\sigma ^au  \label{76}
\end{equation}
\begin{equation}
L=ik\bar \zeta \sigma ^{-2a}  \label{77}
\end{equation}
It is very similar to the Kramer solution for $a=1$. For $a=1/2$ , $P$ is
still given by Eq. (76) while $L=ik\bar \zeta \sigma ^{-1}\ln \sigma $. The
energy density is 
\begin{equation}
n^2=\frac{6m_0}u-2\delta _{1,2a}u^2  \label{78}
\end{equation}

\section{Solutions with exponential behaviour}

These are solutions with $P=pe^{cu}$ and $C=-c^2$ in Eq. (46) or Eqs. (52),
(53) and (54). We shall give two solutions with y-symmetry. First we take $%
q_x=2\left( 1-b\right) x^{b-1}$, where the constant $b\neq 1$ and put this
expression in Eq. (53): 
\begin{equation}
x^2p_{xx}+\left( b-2\right) xp_x+\left[ \frac 14\left( b-2\right) \left(
b-3\right) +4c^2\left( 1-b\right) ^2x^{2b}\right] p=0  \label{79}
\end{equation}
When $b=0$ this is the Euler equation, similar to Eq. (64) and has three
types of solutions. The first two of them read 
\begin{equation}
p=\mid x\mid ^{\frac 32\pm \gamma }  \label{80}
\end{equation}
\begin{equation}
p=\mid x\mid ^{\frac 32}\ln \mid x\mid  \label{81}
\end{equation}
when $c^2<3/16$ or $c^2=3/16$ respectively. Here $\gamma =\frac 12\left(
3-16c^2\right) ^{1/2}$. When $b\neq 0$ the solution is given by Bessel
functions: 
\begin{equation}
p=x^{\frac{3-b}2}\left[ C_1J_\nu \left( \varepsilon x^b\right) +C_2Y_\nu
\left( \varepsilon x^b\right) \right]  \label{82}
\end{equation}
where $\nu =\sqrt{3-b}/2b,$ $\varepsilon =2\mid b-1\mid c/b$. In all cases 
\begin{equation}
L=i\sqrt{2}\left( 1-b\right) x^{b-1}  \label{83}
\end{equation}
These are essentially the solutions found in [4] where x-symmetry was used
instead together with a special transformation between a $u$-independent and
a $u$-dependent solution.

The second solution to be presented is an analog of Eq. (69) i.e. $P=e^{cu}$
depends only on $u$. Then Eq. (53) may be integrated to 
\begin{equation}
Q=\frac 12\left( b^2-\frac 23c^2q_x^4+2dq_x\right) ^{1/2}  \label{84}
\end{equation}
where $b>0$ and $d$ are constants. When $d=0$ Eq. (84) is solved by elliptic
functions: 
\begin{equation}
L=i\frac b\lambda cn\left( \lambda x\right)  \label{85}
\end{equation}
where $\lambda =\left( \frac 83b^2c^2\right) ^{1/4}$ and the modulus of $cn$
is $1/2.$ It is clear from Eq. (84) that $Q\leq b/2$. Eq. (55) yields 
\begin{equation}
n^2=6cm_0-4c^2Q^2e^{4cu}  \label{86}
\end{equation}
The second term is definitely negative and if $m_0>0,$ $c>0$ dominates for
large $u$, no matter how big $m_0$ is. However, if $m_0<0,$ $c<0$ the first
term remains positive while the second is damped exponentially for large
retarded times. Choosing $\mid m_0\mid $ big enough we can arrange for $%
n^2>0 $.

Let us give next an example of an axisymmetric solution. Let us substitute
in Eq. (54) $q_\sigma =Q=a=const$ to obtain the generalization of Eq. (72): 
\begin{equation}
p_{\sigma \sigma }+\frac 1\sigma p_\sigma +a^2c^2p=0  \label{87}
\end{equation}
It is solved by Bessel functions e.g. 
\begin{equation}
P=e^{cu}J_0\left( ac\sigma \right)  \label{88}
\end{equation}
\begin{equation}
L=ia\bar \zeta  \label{89}
\end{equation}
This solution was found and discussed in [9].

Finally, we give an example of a solution without y-symmetry or axial
symmetry. We shall obtain the most general KS pure radiation field with $%
P=pe^{cu}$. The results of Section IV and Eq. (56) show that 
\begin{equation}
q=\frac 1{2ic}\ln \frac B{\bar B}  \label{90}
\end{equation}
\begin{equation}
P=\left( B\bar B\right) ^{1/2}e^{cu}  \label{91}
\end{equation}
where $B$ satisfies $\psi =0$, namely 
\begin{equation}
B^2\left( \ln B\right) _{\zeta \bar \zeta }=S\left( \zeta \right)  \label{92}
\end{equation}
$S$ is an arbitrary analytic function. If $S\neq 0$ we introduce the new
variable $\omega =\int^\zeta S\left( \zeta ^{\prime }\right) d\zeta ^{\prime
}$ and Eq. (92) transforms into the Liouville equation 
\begin{equation}
B^2\left( \ln B\right) _{\omega \bar \zeta }=1  \label{93}
\end{equation}
Returning to the original variables, its general solution reads 
\begin{equation}
B=S\left( \zeta \right) ^{1/2}\frac{f\left( \zeta \right) +g\left( \bar \zeta
\right) }{\sqrt{f\left( \zeta \right) _\zeta g\left( \bar \zeta \right) _{%
\bar \zeta }}}  \label{94}
\end{equation}
The functions $f$ and $g$ are arbitrary. When $S=0$ the solution of Eq. (92)
is 
\begin{equation}
B=f\left( \zeta \right) g\left( \bar \zeta \right)  \label{95}
\end{equation}
It has zero twist i.e. it is a RT solution. Then we may set $L=q=0$ which
means $f\left( \zeta \right) \equiv g\left( \zeta \right) $ and $P=f\bar f$ $%
e^{cu}$. A coordinate transformation sets $f=1$. Therefore the solution (95)
is equivalent to $B=1$.

In [3,15] the general axisymmetric pure radiation KS fields were found.
Interestingly enough, their $u$-dependence covers the three types (49), (50)
and (51) with $A=0$. The exponential solution was generalized in [3] to a
non-symmetric one with 
\begin{equation}
B=G\left( \zeta \right) +\bar \zeta K\left( \zeta \right)  \label{96}
\end{equation}
$G$ and $K$ being arbitrary. Eq. (96) may be obtained from the general
formula (94) by specializing to $g\left( \bar \zeta \right) =\bar \zeta $: 
\begin{equation}
B=f\left( \zeta \right) \left[ \frac{S\left( \zeta \right) }{f\left( \zeta
\right) _\zeta }\right] ^{1/2}+\bar \zeta \left[ \frac{S\left( \zeta \right) 
}{f\left( \zeta \right) _\zeta }\right] ^{1/2}  \label{97}
\end{equation}

The general KS field with non-radiative Maxwell field has been given in
[11]. It was remarked in [1] that no solution with a null Maxwell field has
been found. Formulae (94) and (95) allow the study of this question in the
context of solutions with separated variables. The energy density (63)
tolerates exponential dependence. Eq. (55) reads in this case: 
\begin{equation}
3m_0c=\kappa N\left( \zeta \right) \bar N\left( \bar \zeta \right) B\bar B
\label{98}
\end{equation}
Obviously $B$ can't be given by Eq. (94) but a non-twisting solution with $%
B=1$ is possible. Then $P=e^{cu}$ and $\Phi _2^0=\left( 3m_0c/\kappa \right)
^{1/2}$. The mass parameter is constant, as usual, $m=m_0$. This is nothing
but a special case of the Einstein-Maxwell solution of Robinson and Trautman
[10] in disguise (see also Eq. (24.41) in [1]). The conclusion is that there
is no twisting KS solution with a null Maxwell field and separated variables.

\section{Conclusions}

In this paper we have carried out the idea of Stephani based on the
introduction of an invariant potential for algebraically special, twisting
and expanding gravitational fields. It works best in the important subclass
of fields with vanishing NUT parameter. The system of Einstein-Maxwell
equations is reduced to the couple of equations (25) and (26) for $P$ and $L$
where $L$ is represented by a real function of two variables $q\left( \zeta ,%
\bar \zeta \right) $ (the $L_u=0$ gauge). We have studied mainly pure
radiational solutions for which Eq. (26) is an inequality. The main equation
(25), originally of fourth order, becomes a linear second order equation for 
$P$ (39), which further simplifies for the most common symmetries. In some
cases it is linear for $q_x$ too. Of course, it is not possible to enumerate
all solutions of Eq. (39). The method of separation of variables was used to
systematize most of the known solutions obtained in the past by a variety of
different approaches. We have also investigated the region where the energy
density of pure radiation is positive. Samples of new solutions were given
to illustrate the application of Eqs. (39), (53), (54) and (55). Three types
of behaviour with respect to the retarded time have been found: exponential,
trigonometric and linear. Among the new solutions is the most general KS
pure radiation field with exponential time behaviour. It was shown that the
radiation field cannot be a null Maxwell field unless the twist vanishes.
Finally, it should be mentioned that the case $m=0,$ $M\neq 0$ is much more
difficult because Eq. (9) remains non-linear in $P$.

\section*{Acknowledgments}

This work was supported by the Bulgarian National Fund for Scientific
Research under Contract No.F-632.


\begin{references}
\bibitem{one}  D. Kramer, H. Stephani, M. MacCallum and E. Herlt, {\it Exact
Solutions of Einstein's Field Equations }(Cambridge University Press,
Cambridge, 1980).

\bibitem{two}  E. Herlt and H. Stephani, Class. Quantum Grav. {\bf 1}, 95
(1984)

\bibitem{three}  H. Stephani, Gen. Rel. Grav. {\bf 15}, 173 (1983)

\bibitem{four}  J. Tafel, P. Nurowski and J. Lewandowski, Class. Quantum
Grav. {\bf 8}, L83 (1991)

\bibitem{five}  J. Lewandowski, P. Nurowski and J. Tafel, Class. Quantum
Grav. {\bf 8}, 493 (1991)

\bibitem{six}  A. Grundland and J. Tafel, Class. Quantum Grav. {\bf 10},
2337 (1993)

\bibitem{seven}  D. Kramer and U. H\"ahner, Class.Quantum Grav. {\bf 12},
2287 (1995)

\bibitem{eight}  U. von der G\"onna and D. Kramer, Class.Quantum Grav. {\bf %
15}, 2017 (1998)

\bibitem{nine}  H. Stephani, J. Phys. A {\bf 12}, 1045 (1979)

\bibitem{ten}  I. Robinson and A. Trautman, Proc. Royal Soc. London A {\bf %
265}, 463 (1962)

\bibitem{eleven}  G. C. Debney, R. P. Kerr and A. Schild, J. Math. Phys. 
{\bf 10}, 1842 (1969)

\bibitem{twelve}  D. W. Trim and J. Wainwright, J. Math. Phys. {\bf 15}, 535
(1974)

\bibitem{thirteen}  A. D. Rendall, Class.Quantum Grav. {\bf 5}, 1339 (1988)

\bibitem{fourteen}  D. Singleton, Class.Quantum Grav. {\bf 7}, 1330 (1990)

\bibitem{fifteen}  E. Herlt, Gen. Rel. Grav. {\bf 12}, 1 (1980)
\end{references}
\end{document}